\documentclass[apj]{emulateapj}

\newcommand{\ms}{\mbox{m\,s$^{-1}$}}

\newcommand{\Msun}{\mbox{M$_{\odot}$}}
\newcommand{\Rsun}{\mbox{R$_{\odot}$}}
\newcommand{\Mjup}{\mbox{M$_{\rm Jup}$}}
\newcommand{\Rjup}{\mbox{R$_{\rm Jup}$}}

\slugcomment{}

\shorttitle{Precise Radial Velocity Measurements for  {\it Kepler} Giants}
\shortauthors{Sato et al.}

\begin{document}


\title{Precise Radial Velocity Measurements for {\it Kepler} Giants
Hosting Planetary Candidates: Kepler-91 and KOI-1894}


\author{Bun'ei Sato\altaffilmark{1},
Teruyuki Hirano\altaffilmark{1},
Masashi Omiya\altaffilmark{2},
Hiroki Harakawa\altaffilmark{2},
Atsushi Kobayashi\altaffilmark{1},
Ryo Hasegawa\altaffilmark{1},
Takuya Takarada\altaffilmark{1},
Kiyoe Kawauchi\altaffilmark{1},
and Kento Masuda\altaffilmark{3}
}
\email{satobn@geo.titech.ac.jp}

\altaffiltext{1}{Department of Earth and Planetary Sciences, Tokyo Institute of
Technology, 2-12-1 Ookayama, Meguro-ku, Tokyo 152-8551, Japan}
\altaffiltext{2}{National Astronomical Observatory of Japan, 2-21-1 Osawa,
   Mitaka, Tokyo 181-8588, Japan}
\altaffiltext{3}{Department of Physics, The University of Tokyo, Tokyo
113-0033, Japan}

%% \altaffiltext{5}{Patron, Alonso's Bar and Grill}


\begin{abstract}
We present results of radial-velocity follow-up observations for the two {\it Kepler}
evolved stars Kepler-91 (KOI-2133) and KOI-1894, which had been announced as
candidates to host transiting giant planets, with the Subaru 8.2m telescope and
the High Dispersion Spectrograph (HDS). By global modeling of the
high-precision radial-velocity data taken with Subaru/HDS and photometric ones
taken by {\it Kepler} mission taking account of orbital brightness modulations
(ellipsoidal variations, reflected/emitted light, etc.) of the host stars,
we independently confirmed that Kepler-91 hosts a transiting planet with a mass
of $0.66\Mjup$ (Kepler-91b), and newly detected an offset of $\sim$20 m s$^{-1}$
between the radial velocities taken at $\sim1$-yr interval, suggesting the existence
of additional companion in the system.
As for KOI-1894, we detected possible phased variations in the radial velocities and
light curves with 2--3$\sigma$ confidence level which could be explained
as a reflex motion and ellipsoidal variation of the star caused by the transiting
sub-saturn-mass ($\sim0.18\Mjup$) planet.
\end{abstract}





\section{Introduction}\label{intro}
Detecting planetary transits is highly valuable not only because it makes an
independent confirmation of a planet from radial-velocity observations but also
because the photometric transits provide unambiguous information
on planet masses and radii, and thus mean density and interior structure
\citep[e.g.][and references therein]{winn:2011}.
Various high-precision follow-up studies for transiting planets can also uncover
planetary atmospheres and the (mis-)alignment between the stellar spin and
planetary orbital axes via the Rossiter-McLaughlin effect
\citep[e.g.][]{seager:2010,winn:2011}. These properties
of transiting planets provide valuable hints for planet formation and evolution
processes including dynamical interaction between planets.

Although transiting planets are as such important, it is difficult to detect them
around evolved stars, especially giant stars, due to the large sizes of the host stars.
Targets for on-going radial-velocity surveys for planets around giants have typical radii of
$\sim$10 $\Rsun$ \citep[e.g.,][]{dasilva:2006, takeda:2008, liu:2010, wang:2011,ziel:2012},
and then the relative flux variation of such a
giant host star caused by a transit of a Jupiter-sized planet is only $\sim1\times10^{-4}$,
which is comparable to that for transit of an Earth-sized planet across a solar-type star.
It is impossible to detect such a transit from ground-based photometry and
thus no transiting planets had been found around giant stars.
Our understanding of properties of planets around such stars are thus far
behind from those around solar-like stars, although tens of planets have
been found around giants by precise radial-velocity surveys
\cite[e.g.][and references therein]{sato:2013}.

The {\it Kepler} mission was successful in detecting
transiting planets with very high photometric precision ($\sim2\times10^{-5}$)
since 2009\footnote{NASA Exoplanet Archive (http://exoplanetarchive.ipac.caltech.edu/)}.
To date it has discovered more than 4000 transiting planet candidates from sub-Earth-size
to super-Jupiter-size in $\sim$0.3--2000 d orbits \citep[e.g.][]{batalha:2013}.
Although most of them are orbiting solar-type stars, {\it Kepler}'s photometric precision
is high enough to detect a Jupiter-sized planet transiting a giant star.

As expected, {\it Kepler} has identified several planet candidates around giant stars
with radii larger than $5~R_{\odot}$ \citep{batalha:2013}. The planet candidates
have radii comparable to or larger than Jupiter's, and it is interesting that many of
them have short-period orbits. Since such short-period planets have rarely been
found by radial-velocity surveys around evolved stars \citep[e.g.][]{johnson:2007, sato:2008,
jones:2014},
they will provide us unique opportunities to investigate planet formation and
evolution processes around such stars.

Here we report the results of radial-velocity follow-up observations using Subaru 8.2m
telescope for two of the {\it Kepler} evolved stars, Kepler-91 (KOI-2133) and KOI-1894.
Thanks to the high precision in their radial-velocity measurements, we independently
confirmed a jovian planet (Kepler-91b, KOI-2133.01) previously reported around
Kepler-91 \citep{lillobox:2014a,lillobox:2014b,barclay:2014} and newly found a
hint for the existence of additional companion in the system. We also detected a
possible sub-saturn-mass planet around KOI-1894 (KOI-1894.01) with 2--3$\sigma$
confidence level.

The rest of the paper is organized as follows.
The adopted stellar parameters for the two stars are presented in section \ref{targets}.
The observations are described in section \ref{obs} and the results of global analysis for
the radial velocities and light curves are presented in section \ref{analysis}.
Section \ref{summary} is devoted to discussion and summary.

\section{Targets}\label{targets}
Kepler-91 (KIC 8219268, KOI-2133; $K_p$=12.495\footnote{{\it Kepler} mag}) was
identified as a candidate star hosting a transiting short-period jupiter-sized planet
(KOI-2133.01; $R_p=18.24~R_{\oplus}$, $P=6.2465798\pm0.0000821$ d) by \cite{batalha:2013}.
After that, \cite{lillobox:2014a} reported the confirmation of its planetary nature 
based on the detailed analysis of orbital brightness modulation seen in the light curve caused by
ellipsoidal variation, Doppler boosting, and reflected/emitted light from planet
\citep[e.g.][]{faigler:2011, mazeh:2012}, and then the planet was
named Kepler-91b with the radius $R_p=1.384^{+0.011}_{-0.054}~\Rjup$ and
mass $M_p=0.88^{+0.17}_{-0.33}~\Mjup$.
Although \cite{esteves:2013} and \cite{sliski:2014} claimed a possible non-planertary
nature for the system, \cite{lillobox:2014b} and \cite{barclay:2014} very recently
reconfirmed the planetary nature via radial-velocity measurements with a precision
of $\sim$100 m s$^{-1}$ and $\sim$20 m s$^{-1}$, respectively, and obtained the
planetary mass of 1.09$\pm$0.20$\Mjup$ and 0.73$\pm$0.13$\Mjup$, respectively.

The stellar parameters (effective temperature $T_{\rm eff}$, surface gravity $\log g$,
mass $M_{\star}$, and radius $R_{\star}$) of Kepler-91 were reported by \cite{batalha:2013} to be
$T_{\rm eff}=4712$ K,
$\log g=2.85$ cgs,
$M_{\star}=2.25~M_{\odot}$,
and
$R_{\star}=9.30~R_{\odot}$.
After that, the parameters have been updated by the spectroscopic and asteroseismic
analyses in \cite{huber:2013a} and \cite{lillobox:2014a}, which are consistent with each
other. Thus, we here adopted the values listed in \cite{lillobox:2014a}:
$T_{\rm eff}=4550\pm75$ K,
$\log g=2.953\pm0.007$ cgs,
$M_{\star}=1.31\pm0.10~M_{\odot}$,
and
$R_{\star}=6.30\pm0.16~R_{\odot}$.

KOI-1894 (KIC 11673802; $K_p$=13.427) was also reported to be a planet-host candidate
having a transiting short-period jupiter-sized planet (KOI-1894.01; $R_p=16.29~R_{\oplus}$,
$P=5.2880157\pm0.0000447$ d)
by \cite{batalha:2013}, though radial-velocity follow-up observations for the star
have not been reported yet.
The stellar parameters of KOI-1894 were derived by \cite{batalha:2013}
to be $T_{\rm eff}=4815$ K,
$\log g=2.87$ cgs,
$M_{\star}=2.02~M_{\odot}$,
and
$R_{\star}=8.62~R_{\odot}$,
and they have been updated by the spectroscopic and asteroseismic
analyses in \cite{huber:2013a} as
$T_{\rm eff}=4992\pm75$ K,
$M_{\star}=1.410\pm0.214~M_{\odot}$,
and
$R_{\star}=3.790\pm0.190~R_{\odot}$. Here we adopted the updated values in this paper.
Recently \cite{law:2014} reported a non-detection of blended stars for KOI-1894,
which could have been physically associated companions and/or responsible for transit false positives
if they were within $\sim$0.15$^{\prime\prime}$--2.5$^{\prime\prime}$ separation and with
magnitude difference up to $\Delta m\simeq 6$, by high-angular-resolution AO imaging.
The parameters for the two stars we adopt here are summarized in Table \ref{tbl:stars}.

%----------------------------------------------------------
\begin{deluxetable}{lrr}
%\tabletypesize{\small}
\tablecaption{Stellar Parameters for Kepler-91 and KOI-1894}
\tablewidth{0pt}
\tablehead{
\colhead{Parameter} & \colhead{Kepler-91} & \colhead{KOI-1894}}
\startdata
\label{tbl:stars}
$T_{\rm eff}$ (K) & 4550$\pm$75\tablenotemark{a} & 4992$\pm$75\tablenotemark{b}\\
$\log g$ (cgs) & 2.953$\pm$0.007\tablenotemark{a} & 2.87\tablenotemark{c}\\
$M_{\star}$ ($M_{\odot}$) & 1.31$\pm$0.10\tablenotemark{a} & 1.410$\pm$0.214\tablenotemark{b}\\
$R_{\star}$ ($R_{\odot}$) & 6.30$\pm$0.16\tablenotemark{a} & 3.790$\pm$0.190\tablenotemark{b}
\enddata
\tablenotetext{a}{\cite{lillobox:2014a}}
\tablenotetext{b}{\cite{huber:2013a}}
\tablenotetext{c}{\cite{batalha:2013}}
\end{deluxetable}
%----------------------------------------------------------

\section{Precise Radial Velocity Measurements with Subaru/HDS}\label{obs}

We obtained high-precision radial-velocity data for the stars with the 8.2m Subaru
telescope and the High Dispersion Spectrograph \citep[HDS;][]{noguchi:2002}
in 2013 and 2014.
We used the setups of StdI2b (2013 June 28-30 and 2014 July 8, 10, 14-16)
and StdI2a (2013 December 6 and 2014 July 12-13), which simultaneously
cover a wavelength region of 3500--6200${\rm \AA}$ and 4900--7600${\rm \AA}$
respectively, the image slicer No.2 \citep[IS\#2;][]{tajitsu:2012} yielding
a spectral resolution ($R=\lambda/\Delta\lambda$) of 80000, and an iodine
absorption cell \citep[I$_2$ cell;][]{kambe:2002} for precise radial-velocity measurements.
We obtained a total of 29 and 18 data points for Kepler-91 and KOI-1894
with typical signal-to-noise ratio of S/N$=$30--80 pix$^{-1}$ and 30--45 pix$^{-1}$,
respectively, by an exposure time of 1200 sec depending on weather condition.
The reduction of echelle data (i.e. bias subtraction, flat-fielding,
scattered-light subtraction, and spectrum extraction) was performed
using the IRAF\footnote{IRAF is distributed by the National
Optical Astronomy Observatories, which is operated by the
Association of Universities for Research in Astronomy, Inc. under
cooperative agreement with the National Science Foundation,
USA.} software package in the standard manner.

We performed radial-velocity analysis for I$_2$-superposed stellar spectra
(star+I$_2$) by the method described in \citet{sato:2002} and
\citet{sato:2012}, which is based on the method by
\citet{butler:1996} and \citet{val:95}. A star+I$_2$ spectrum is modeled
as a product of a high resolution I$_2$ and a stellar template spectrum convolved
with a modeled instrumental profile (IP) of the spectrograph.
We obtain the stellar template spectrum by deconvolving a pure stellar
spectrum with the IP estimated from an I$_2$-superposed
Flat spectrum. We achieved a radial-velocity precision of 5--17 m s$^{-1}$ for
Kepler-91 and 9--15 m s$^{-1}$ for KOI-1894.
The derived radial velocities are listed in Table \ref{tbl:rv-koi2133} and Table \ref{tbl:rv-koi1894}
together with the estimated uncertainties, and are plotted in Figure \ref{fig:rv_koi2133_notrend}
and Figure \ref{fig:rv_koi1894}.

%----------------------------------------------------------
\begin{deluxetable}{lrr}
%\tabletypesize{\scriptsize}
\tablecolumns{3}
\tablewidth{0pt}
\tablecaption{Radial Velocities for Kepler-91}
\tablehead{
\colhead{BJD$-$2450000} & \colhead{Velocity (\ms)} & \colhead{Uncertainty
(\ms)}}
\startdata
\label{tbl:rv-koi2133}
6473.00247 & 43.20 & 4.61\\ 
6473.01719 & 33.85 & 4.58\\ 
6473.03192 & 45.26 & 5.03\\ 
6473.96075 & 17.35 & 7.36\\ 
6473.97549 & 8.93 & 7.34\\ 
6473.99022 & 30.04 & 6.39\\ 
6474.95764 & $-$24.28 & 5.43\\ 
6474.97235 & $-$22.62 & 5.03\\ 
6474.98708 & $-$39.46 & 5.50\\ 
6633.69753 & $-$42.12 & 17.38\\ 
6847.93547 & 42.30 & 10.56\\ 
6847.95020 & 37.46 & 13.63\\ 
6847.96513 & 11.49 & 8.76\\ 
6847.98006 & 29.77 & 7.96\\ 
6847.99479 & 17.07 & 6.12\\ 
6848.00952 & 19.94 & 8.61\\ 
6848.08121 & 23.25 & 6.59\\ 
6849.75412 & $-$52.12 & 5.23\\ 
6849.76886 & $-$36.10 & 5.06\\ 
6849.78360 & $-$34.47 & 5.48\\ 
6853.04854 & $-$40.23 & 13.48\\ 
6853.97273 & 5.73 & 5.72\\ 
6853.98746 & 9.02 & 4.73\\ 
6854.00219 & 7.66 & 5.17\\ 
6854.84234 & 2.14 & 4.99\\ 
6854.85708 & 0.36 & 5.25\\ 
6854.87181 & $-$6.69 & 6.03\\ 
6855.80523 & $-$43.37 & 6.63\\ 
6855.81828 & $-$41.91 & 5.28
\enddata
\end{deluxetable}

%----------------------------------------------------------
\begin{deluxetable}{lrr}
%\tabletypesize{\scriptsize}
\tablecolumns{3}
\tablewidth{0pt}
\tablecaption{Radial Velocities for KOI-1894}
\tablehead{
\colhead{BJD$-$2450000} & \colhead{Velocity (\ms)} & \colhead{Uncertainty
(\ms)}}
\startdata
\label{tbl:rv-koi1894}
6473.04710 & $-$9.95 & 9.45\\ 
6473.06182 & $-$0.82 & 11.01\\ 
6473.07656 & $-$11.81 & 9.94\\ 
6474.00523 & $-$5.31 & 10.97\\ 
6474.01995 & 1.95 & 10.25\\ 
6474.03468 & 17.33 & 10.36\\ 
6474.04941 & 5.73 & 10.49\\ 
6475.00210 & $-$3.24 & 10.03\\ 
6475.01683 & $-$21.07 & 9.68\\ 
6475.03159 & $-$13.24 & 8.79\\ 
6475.04632 & $-$6.92 & 9.62\\ 
6848.09625 & $-$4.14 & 15.02\\ 
6849.94269 & 0.52 & 10.34\\ 
6849.95743 & 13.69 & 9.86\\ 
6849.97216 & 10.09 & 9.04\\ 
6855.88939 & 0.65 & 9.63\\ 
6855.90411 & $-$3.06 & 8.95\\ 
6855.91885 & 6.48 & 9.09
\enddata
\end{deluxetable}
%----------------------------------------------------------

\section{Radial Velocity and Light Curve Analysis}\label{analysis}
\subsection{Method and Light Curve Reduction}\label{method}
The variations in the observed radial velocities for both targets show a sign of planetary
companions to Kepler-91 and KOI-1894. In order to obtain accurate and precise estimates
for system parameters of those systems, 
we here present a global analysis that makes use of all the available information 
from the {\it Kepler} photometry and our spectroscopy.  

As is well known, a very precise light curve of a star orbited by planet(s)
shows a periodic modulation due to several astrophysical effects: the ellipsoidal 
variation, Doppler boosting, and reflected/emission light from the planet 
\citep[e.g.,][]{faigler:2011, mazeh:2012}. 
The last two effects (Doppler boosting and planetary reflection/emission) are synchronous 
with the planet's orbital period $P_\mathrm{orb}$, with different peak locations
along the planet phase ($\phi\sim 0.25$ for boosting and $\phi\sim0.5$ for
the planetary reflection/emission, respectively). On the other hand, the ellipsoidal variation,
which is caused by a tidal distortion by the planet's gravity, have two flux peaks
($\phi\sim 0.25$ and 0.75) within one orbit. Thanks to the different phase
dependence of these three effects, a very precise light curve and its modeling enable
us to distinguish these three effects, and we can extract physical parameters
such as the planet-to-star mass ratio $q$ and scaled semi-major axis $a/R_{\star}$,
where $a$ is semi-major axis and $R_{\star}$ is stellar radius. 
Moreover, for transiting systems as in the present cases, incorporating 
the transit and/or secondary eclipse model into the above 
phase-curve variation lets us learn more about the planet properties
(e.g., planet-to-star radius ratio $R_p/R_{\star}$ and orbital inclination $i_o$). 

To obtain the phase-folded light curves for Kepler-91 and KOI-1894,
and estimate system parameters, we downloaded all available public 
light curves (Q0--Q17) from {\it Kepler} MAST archive. While only long-cadence data were 
available for Kepler-91, KOI-1894's light curves involve some short-cadence 
data. Adopting the PDC-SAP flux data, for which unphysical artifacts are detrended, 
we reduced the light curves by the following procedure.
First, after removing planetary transits, we further detrended and normalized the light curve 
for each quarter by fitting it with a fifth-order polynomial so as to remove the long-term trends 
that were not removed in the PDC-SAP flux\footnote{Note that since the 
orbital periods of our targets are much shorter than the time span of one 
quarter, the flux modulations due to ellipsoidal variations, Doppler boosting, 
and planetary reflection are retained in this process.}. 
This process was repeated implementing a $5\sigma$ clipping to remove outliers. 
We then combined the light curves for all quarters and phase-folded them with 
the ephemerides derived by the official {\it Kepler} team using the Q1--Q16 data;
orbital period $P_\mathrm{orb}=6.24668005~(\pm0.00002647)$ d and
transit center $T_c^{(0)}=2454969.38661~(\pm 0.00346)$ BJD for Kepler-91,
and $P_\mathrm{orb}=5.28789787~(\pm0.00001198)$ d and
$T_c^{(0)}=2454968.36353~(\pm 0.00189)$ BJD for KOI-1894 (from NASA Exoplanet Archive).
We later consider the impact of incorrect ephemeris. Finally, the folded light curves
were binned into 300 and 250 phase bins for Kepler-91 and KOI-1894,
respectively. These bin numbers were adopted so that each bin approximately
covers a cycle span of long-cadence data ($\sim 30$ minutes). 
The flux error for each bin $\sigma_\mathrm{LC}$, which is the standard deviation
of the mean flux, was computed based on the dispersion of the
flux values within the bin.
The long-cadence and short-cadence flux data were separately folded and 
binned for KOI-1894, but the binned light curve of the short-cadence data
was much noisier than that for long-cadence. We thus decided to ignore the
short-cadence data in the following analysis.

The analysis below is based on the method by \cite{hirano:2015},
who employed the EVIL-MC model \citep{jackson:2012} for the phase-curve variation
(i.e., ellipsoidal variations, Doppler boosting, and planetary reflection) 
with some revisions (e.g., application to an eccentric orbit). 
This phase-curve model is multiplied by the analytic transit model 
by \citet{ohta:2009}, and the relevant parameters (e.g., $a/R_{\star}$ and $i_o$) are 
simultaneously determined. In addition, we also model and fit the observed 
radial velocities $v_\mathrm{obs}$. Thus, the $\chi^2$ statistics in the
present case is
%%%%%%%%%%%%%%%%%
\begin{eqnarray}
\label{eq:chisq}
\chi^2 =\sum_{i}\frac{(v_\mathrm{obs}^{(i)}-v_\mathrm{model}^{(i)})^2}{\sigma_\mathrm{RV}^{(i)2}}
+\sum_{i}\frac{(f_\mathrm{obs}^{(i)}-f_\mathrm{model}^{(i)})^2}{\sigma_\mathrm{LC}^{(i)2}},
\end{eqnarray}
%%%%%%%%%%%%%%%%%
where $v_\mathrm{obs}^{(i)}$ and $\sigma_\mathrm{RV}^{(i)}$ 
are the $i$-th observed radial velocity and its error, and 
$f_\mathrm{obs}^{(i)}$ and $\sigma_\mathrm{LC}^{(i)}$ 
are the $i$-th observed light curve flux and its error, which are the mean flux
and its standard deviation in the $i$-th bin, respectively \citep[e.g., ][]{hirano:2011}. 
The radial velocity is modeled as
%%%%%%%%%%%%%%%%%
\begin{eqnarray}
\label{eq:RVmodel}
v_\mathrm{model}=K\{\cos(f+\omega)+e\cos\omega\}+\gamma, 
\end{eqnarray}
%%%%%%%%%%%%%%%%%
where $K$, $f$, $e$, $\omega$, and $\gamma$ are the radial velocity semi-amplitude, 
true anomaly, orbital eccentricity, argument of periastron, and
radial velocity offset of our dataset, respectively. 
For the light curve model $f_\mathrm{model}$, we refer the readers to \cite{hirano:2015}, 
for details. In the next subsections, we describe the fitting 
procedure for each of the targets.

\subsection{Kepler-91}
\begin{figure}
\epsscale{1.2}
\plotone{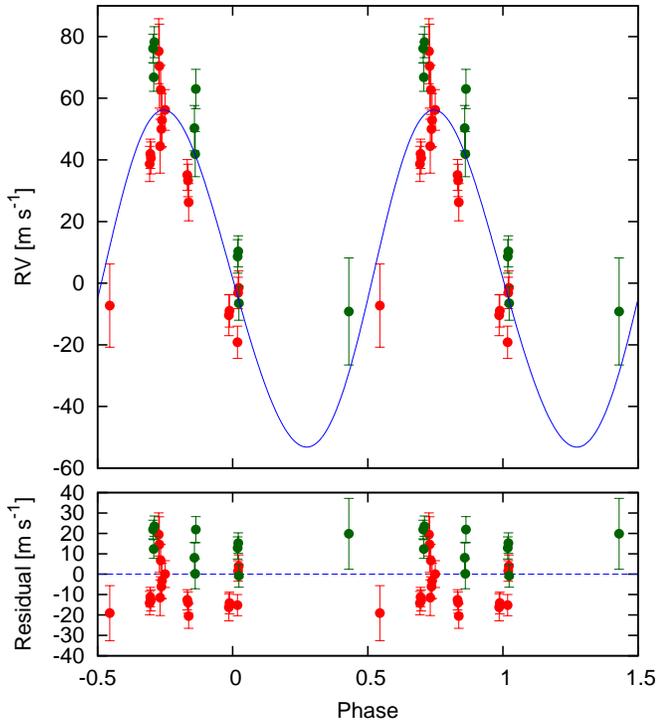}
\caption{Phased radial velocities for Kepler-91 in the absence of radial-velocity
trend $\dot{\gamma}$.
The data in 2013 and 2014 are shown by green and red points, respectively. The solid
line is the best-fit radial-velocity curve to the data obtained by global modeling of
photometric and radial-velocity data without considering $\dot{\gamma}$.}
\label{fig:rv_koi2133_notrend}
\end{figure}

\begin{figure}
\epsscale{1.2}
\plotone{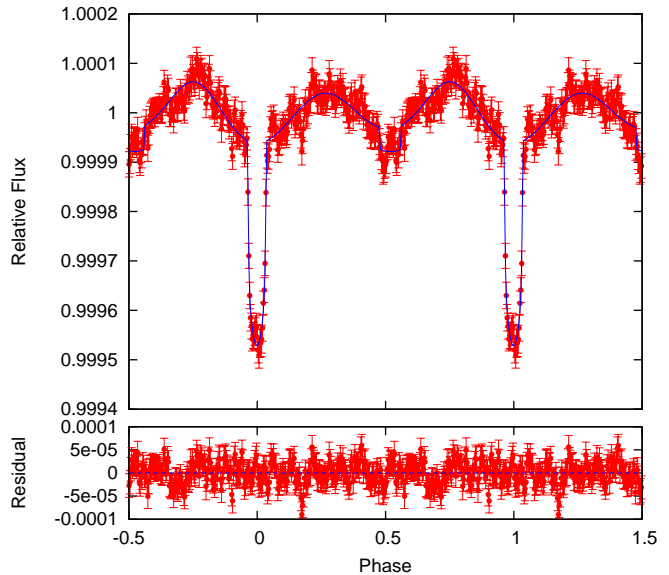}
\caption{Phased light curve for Kepler-91 (red points). The solid line is the best-fit
light-curve model to the data obtained by global modeling of photometric and
radial-velocity data considering the velocity trend ($\dot{\gamma}=-0.0612$ m s$^{-1}$day$^{-1}$).}
\label{fig:phase_koi2133}
\end{figure}

The binned light curve of Kepler-91 in Figure \ref{fig:phase_koi2133} shows a clear pattern of phase-curve
variation; the double peaks at $\phi\sim 0.25$ and 0.75 are representative
of the ellipsoidal variation. In addition, Figure \ref{fig:phase_koi2133} suggests a possible 
detection of secondary eclipse at $\phi\sim 0.5$, which means the reflected/emitted light
from the planet is visible in the folded light curve. Therefore, following 
\citet{jackson:2012} and \cite{hirano:2015}, we model the planet's light $F_p$ by the following
expression:
\begin{eqnarray}
F_p = F_0 - F_1\left(\frac{1+e\cos f}{1-e\sin\omega}\right)^2\sin(f+\omega-2\pi\Delta\phi),
\end{eqnarray}
where $F_1$ is the flux variation amplitude of planet's reflected/emitted light and 
$F_0$ is the planetary flux offset arising from the homogeneous surface emission. 
Following \citet{demory:2013}, \citet{esteves:2014}, and \citet{faigler:2014},
we here introduce the ``phase-delay" $\Delta\phi$ of the brightest part on the planet surface
from the substellar point. The planetary light $F_p$ is added to the flux model
for the beaming and ellipsoidal variation, integrated over the stellar visible hemisphere
and wavelength through the {\it Kepler} band. 
In integrating the local flux calculated from the EVIL-MC model, we assume
that the stellar effective temperature is 4550 K, and adopt the gravity darkening
exponent of $\beta=0.093$ for Equation (10) of \citet{jackson:2012} based on
the theoretical calculation by \citet{claret:1998}. 
The free parameters relevant to the light curve model are
$a/R_{\star}$, the transit impact parameter $b$, limb-darkening coefficients $u_1+u_2$ and $u_1-u_2$
for the quadratic limb-darkening law, $q$ ($=M_p/M_{\star}$), $F_0$, $F_1$, $\Delta\phi$, 
the overall normalization factor $C$ for the folded light curve, $R_p/R_{\star}$, $e\cos\omega$, 
$e\sin\omega$, and $\Delta T_c$, which represents the small time deviation of the transit center from
the ephemeris reported by the {\it Kepler} team.
In our updated ephemeris, the initial transit center becomes $T_c^{(0)}-\Delta T_c$.
We fix the orbital period to be $P_\mathrm{orb}=6.24668005$ d that are
derived by the {\it Kepler} team based on the Q1--Q16 data (see section \ref{method}).

%----------------------------------------------------------
%----------------------------------------------------------

Assuming that the likelihood is proportional to $\exp(-\chi^2/2)$ in Equation 
(\ref{eq:chisq}), we simultaneously model the observed radial velocity and
light curve, and compute the posterior distribution for each fitting parameter by
implementing Markov Chain Monte Carlo (MCMC) simulation. 
In addition to the above mentioned twelve parameters, we add $\gamma$ in Equation
(\ref{eq:RVmodel}) to the fitting parameters. Note that the radial velocity
semi-amplitude $K$ in Equation (\ref{eq:RVmodel}) is related to the planet-to-star
mass ratio $q$ by
%%%%%%%%%%%%%%%%%
\begin{eqnarray}
K=212908.30\left(\frac{M_{\star}/M_\odot}{P_\mathrm{orb}/\mathrm{day}}\right)^{\frac{1}{3}}
\frac{q}{(1+q)^{\frac{2}{3}}}\frac{\sin i_o}{\sqrt{1-e^2}}~~~~(\mathrm{m~s}^{-1}). 
\end{eqnarray}
%%%%%%%%%%%%%%%%%
In computing the posteriors, we do not impose priors on the fitting parameters
except for $u_1$ and $u_2$; due to the sparse sampling of the long-cadence data
($\sim 30$ minutes) and quality of the binned light curve, the limb-darkening coefficients
are poorly constrained in the absence of priors, and thus we decide to put Gaussian 
priors on the limb-darkening coefficients based on the theoretical table by \citet{claret:2011}
as $u_1+u_2=0.74\pm 0.01$ and $u_1-u_2=0.49\pm 0.01$. In our MCMC algorithm,
originally developed in \citet{hirano:2012}, the step size of each fitting parameter is iteratively 
scaled so that the overall acceptance ratio falls between 15\% and 35\% . 
After running 1,000,000 chains, the best-fit value and uncertainty
for each fitting parameter are estimated from the median, and 15.87 and 84.13 percentiles
of the marginalized posterior distribution of that parameter. 

Shortly after we performed the first MCMC trial, we noticed a possible trend 
or drift in the observed radial velocities. Figure \ref{fig:rv_koi2133_notrend} plots the best-fit 
radial velocity curve as a function of the orbital phase $\phi$. For clarity, the data taken in 2013
are shown in green and those by the 2014 campaign are plotted in red. 
While we have detected a clear modulation with an amplitude of $\sim 50$ m s$^{-1}$, 
radial velocities take lower values for the 2014 data, 
which is evident in the bottom panel indicating the residual of the observed 
radial velocities from the best-fit model. 

%----------------------------------------------------------
\begin{figure}
\epsscale{1.2}
\plotone{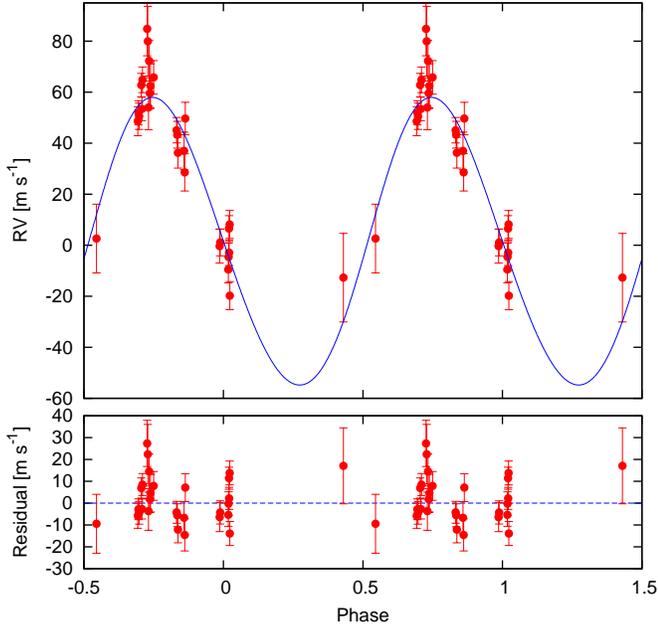}
\caption{Phased radial velocities for Kepler-91 (red points). The solid line is the best-fit radial-velocity curve
to the data obtained by global modeling of photometric and radial-velocity data considering the
velocity trend $\dot{\gamma}$. The derived $\dot{\gamma}$ ($=-0.0612$ m s$^{-1}$day$^{-1}$)
was subtracted from the plot.}
\label{fig:rv_koi2133}
\end{figure}
%----------------------------------------------------------

In order to confirm the presence of the radial velocity drift, we try to fit the data
with an additional parameter: the radial velocity drift $\dot{\gamma}$. 
Adding the drift term $\dot{\gamma}\times t$ to the right side of Equation (\ref{eq:RVmodel}), 
we performed again the MCMC simulation to fit both radial velocity and light curve. 
The derived best-fit parameters and their uncertainties are summarized in 
Table \ref{tbl:best-fit-Kepler91}. To compare between the radial velocity models with and without
a drift term, we compute Bayesian Information Criteria (BIC) for the two cases, which
is computed as $\mathrm{BIC}=\chi^2+n\ln(N_\mathrm{data})$, where $n$ is the number of
free parameters and $N_\mathrm{data}$ is the number of data points. 
In the absence of a trend, we obtain $\mathrm{BIC}=477$, and for the case
of the radial velocity model with $\dot{\gamma}$, BIC becomes 383;
$\Delta\mathrm{BIC}$ is much larger than 10, meaning that the model
with $\dot{\gamma}$ is strongly favored. Therefore, we conclude that
a radial velocity drift (or trend) is present in our dataset, and report the
best-fit parameters for the case with $\dot{\gamma}$ as the final result.
We also tried several periods around $P_\mathrm{orb}$ and found that the
$P_\mathrm{orb}-\sigma$ gave slightly better results in the global fitting than
$P_\mathrm{orb}$ did ($\Delta{\chi^2}\simeq7$). However, the resultant parameters
for the two cases are well consistent within 0.2$\sigma$ level.
\footnote{$\Delta T_c$ differs by $\sim1.5\sigma$ level.}

Figures \ref{fig:phase_koi2133} and \ref{fig:rv_koi2133} plot the folded light curve 
and radial velocities (red points)
along with their best-fit models (blue line). The constant radial velocity offset and 
trend are both removed from the radial velocity data in Figure \ref{fig:rv_koi2133}.
The bottom panel in each figure shows the residuals from the best-fit model.
From the posterior distribution of the fitting parameters, we also estimate
the orbital and planetary parameters (e.g., the orbital inclination $i_o$, planet mass $M_p$
and radius $R_p$) assuming the stellar properties reported by asteroseismology 
(Table \ref{tbl:stars}) to be $i_o=67.37^{+0.63}_{-0.65}$ deg, $M_p=0.66\pm 0.06~\Mjup$,
and $R_p=1.40\pm 0.04~\Rjup$.
These estimates are also shown in Table \ref{tbl:best-fit-Kepler91}. 

%----------------------------------------------------------
\begin{deluxetable}{lrr}
%\tabletypesize{\scriptsize}
\tablecolumns{5}
\tablewidth{0pt}
\tablecaption{Orbital Parameters for Kepler-91}
\tablehead{
\colhead{Parameter} & \colhead{Value (with trend)} & \colhead{Value (without trend)}}
\startdata
\label{tbl:best-fit-Kepler91}
$a/R_{\star}$ & 2.253$^{+0.046}_{-0.045}$ & 2.238$^{+0.046}_{-0.043}$\\
$b$ & 0.9041$^{+0.0061}_{-0.0066}$ &  0.9062$^{+0.0058}_{-0.0062}$\\
$u_1+u_2$ & 0.737$^{+0.010}_{-0.0098}$ &  0.7370$^{+0.0100}_{-0.0098}$\\
$u_1-u_2$ & 0.493$^{+0.010}_{-0.010}$ &  0.493$^{+0.0099}_{-0.010}$\\
$q(=M_p/M_{\star})$ (10$^{-4}$)& 4.82$^{+0.19}_{-0.19}$ &  4.72$^{+0.18}_{-0.18}$\\
$F_0$ (10$^{-5}$) & 5.65$^{+0.57}_{-0.55}$ &  5.63$^{+0.57}_{-0.58}$\\
$F_1$ (10$^{-5}$) & 1.41$^{+0.30}_{-0.29}$ &  1.42$^{+0.30}_{-0.28}$ \\
$\Delta\phi$ & 0.394$^{+0.029}_{-0.033}$ &  0.395$^{+0.028}_{-0.032}$ \\
$R_p/R_{\star}$ & 0.02286$^{+0.00031}_{-0.00030}$ &  0.02294$^{+0.00030}_{-0.00030}$ \\
$e\cos\omega$ & 0.0280$^{+0.0038}_{-0.0042}$ &  0.0277$^{+0.0040}_{-0.0039}$ \\
$e\sin\omega$ & $-$0.043$^{+0.011}_{-0.011}$  & $-$0.046$^{+0.011}_{-0.011}$ \\
$\Delta T_c$ (10$^{-3}$ day)& $-$1.09$^{+0.34}_{-0.34}$ &  $-$1.11$^{+0.34}_{-0.34}$ \\
$\gamma$ (m s$^{-1}$) & $-$42.5$^{+2.0}_{-2.1}$ &   $-$33.2$^{+1.8}_{-1.8}$ \\
$\dot{\gamma}$ (m s$^{-1}$ day$^{-1}$)& $-$0.0612$^{+0.0063}_{-0.0062}$ &    -- \\
$\chi^2$ & 296 & 396\\
BIC & 383 & 477\\
\hline
$i_o$ (deg) & 67.37$^{+0.63}_{-0.65}$ & 67.20$^{+0.65}_{-0.66}$ \\
$e$ & 0.0519$^{+0.0095}_{-0.0088}$ & 0.0535$^{+0.0099}_{-0.0091}$ \\
$\omega$ (deg) & $-$57.2$^{+8.7}_{-7.1}$ & $-$58.8$^{+8.4}_{-6.4}$ \\
$M_p$ ($\Mjup$) & 0.66$\pm$0.06 & 0.65$\pm$0.06 \\
$R_p$ ($\Rjup$) & 1.40$\pm$0.04 & 1.41$\pm$0.04
\enddata
\end{deluxetable}

\subsection{KOI-1894}
The radial-velocity variations are less visible for KOI-1894 owing to the large radial-velocity
error compared to the small semi-amplitude. The phase-folded light curve also shows a very tiny
variation, if any, along the orbital phase. To extract possible planetary signals, 
we simultaneously model the radial velocities and light curve as in the case of Kepler-91. 
Since the observed transit depth and KOI-1894's stellar radius suggest
that the radius of the transiting companion (KOI-1894.01) is no more than $0.65R_J$,
the reflected/emitted light from the planet is expected to be very small. According 
to Equations (1) -- (3) in \citet{shporer:2011}, the planetary reflection is estimated to be $\lesssim 1$ ppm
at the location of KOI-1894.01, which is smaller than the expected
amplitude of the ellipsoidal variation ($\sim 2$ ppm). Visual inspection of the binned
light curve also suggests the absence of secondary eclipse. Thus for simplicity, we here 
neglect the planet flux $F_p$ for KOI-1894, and model the folded light curve with 
the ellipsoidal variation (including Doppler boosting) and transit only. 
Neglecting $F_p$ also helps to avoid $\chi^2$ being
stuck at a local minimum during the optimization. 

%----------------------------------------------------------
\begin{figure}
\epsscale{1.2}
\plotone{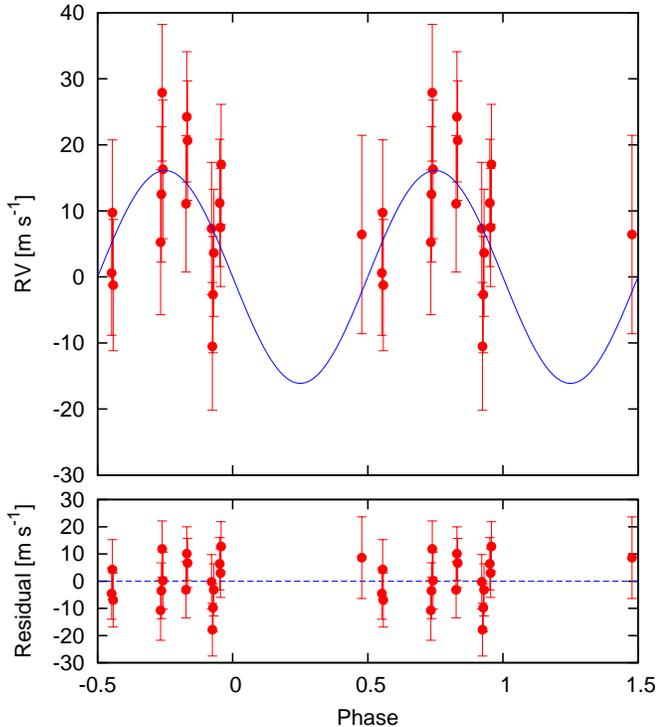}
\caption{Phased radial velocities for KOI-1894 (red points). The solid line is the best-fit radial-velocity curve
to the data obtained by global modeling of photometric and radial-velocity data with the orbital eccentricity
fixed to 0 and the stellar-density prior imposed.}
\label{fig:rv_koi1894}
\end{figure}
%----------------------------------------------------------

Fixing the system parameters as $T_\mathrm{eff}=4992$ K, $M_{\star}=1.41M_\odot$, 
the gravity darkening exponent of $\beta=0.097$ \citep{claret:1998}, and 
$P_\mathrm{orb}=5.28789787~\mathrm{d}$ derived with the Q1--Q16 data
(see section \ref{method}),
we perform MCMC simulations for KOI-1894 to infer the posterior distributions
for the fitting parameters.
In the present case, we have 
the eleven free parameters: $a/R_{\star}$, $b$, $u_1+u_2$, $u_1-u_2$, $q$, $C$, $R_p/R_{\star}$, 
$e\cos\omega$, $e\sin\omega$, $\Delta T_c$, and $\gamma$. 
Again, we assume Gaussian priors for the limb-darkening parameters as 
$u_1+u_2=0.72\pm0.01$, and $u_1-u_2=0.39\pm 0.01$ from the table 
by \citet{claret:2011}.
Due to the weak radial-velocity signal of KOI-1894.01, we found
that the global fit to the radial velocities and light curve exhibits a
degeneracy between the fitting parameters (e.g., $a/R_\star$ and $q$). 
Thus, for KOI-1894, we also impose an additional prior
on the host star's density from Table \ref{tbl:stars} ($\rho_\star=(0.0259\pm0.0055)\rho_\odot$), 
assuming a Gaussian distribution \citep{seager:2003}. 
Otherwise, the fitting algorithm is exactly the same as for Kepler-91. 
The result of the fit is summarized in Table \ref{tbl:best-fit-KOI1894}. The best-fit value
for the mass ratio is $q=(1.02^{+0.44}_{-0.49})\times 10^{-4}$, implying 
$\sim 2\sigma$ detection of the planet. The mass and radius ratios are translated as
$M_p= 0.15^{+0.07}_{-0.08}~\Mjup$ and $R_p=0.64^{+0.04}_{-0.03}~\Rjup$ assuming the stellar
mass and radius in Table \ref{tbl:stars}.
Table \ref{tbl:best-fit-KOI1894} also shows the result of our fit in the absence of
the Gaussian prior on $\rho_\star$. As expected, the mass constraint is
slightly weaker for this case.

%----------------------------------------------------------
\begin{figure}
\epsscale{1.2}
\plotone{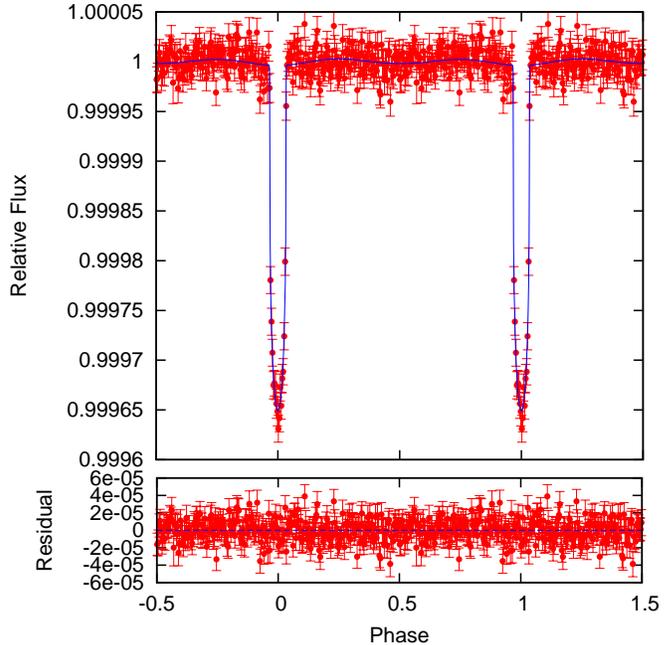}
\caption{Phased light curve for KOI-1894 (red points). The solid line is the best-fit light-curve model
to the data obtained by global modeling of photometric and radial-velocity data with the orbital eccentricity
fixed to 0 and the stellar-density prior imposed.}
\label{fig:phase_koi1894}
\end{figure}
%----------------------------------------------------------

\begin{figure}
\epsscale{1.2}
\plotone{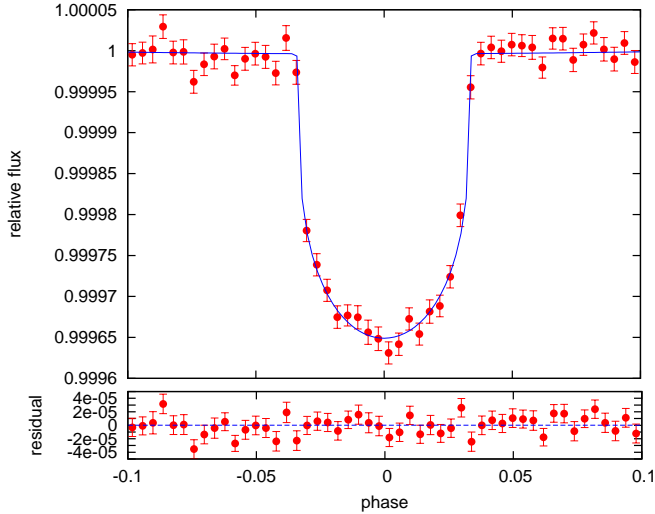}
\caption{Phased light curve for KOI-1894, zoomed in to show transit. The points and line are the
same as those in figure \ref{fig:phase_koi1894}.}
\label{fig:transit_koi1894}
\end{figure}

\begin{figure}
\epsscale{1.2}
\plotone{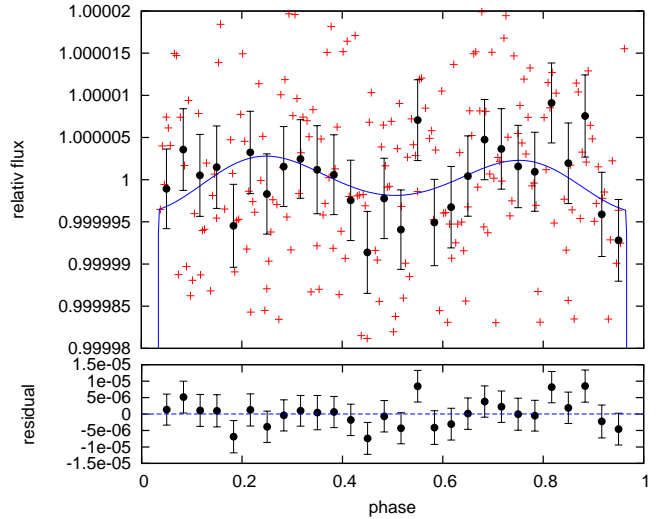}
\caption{Phase-folded light curve for KOI-1894, zoomed in to show 
the flux modulation by ellipsoidal variation. The red crosses are binned flux
data used for the fit and black points with errorbars are data binned into
30 bins. The solid line is the same as that in figure \ref{fig:phase_koi1894}.}
\label{fig:EV_koi1894}
\end{figure}

The orbital eccentricity is poorly constrained due to the small planetary 
signal and small number of radial velocity data, but we can rule out a high eccentricity ($e\gtrsim 0.25$) 
with our current datasets. We also try to fit the data with a circular orbit ($e=0$),
and compare between the eccentric and circular cases.
As shown in Table \ref{tbl:best-fit-KOI1894}, the circular orbital fit
yields a slightly better constraint on the mass ratio; $q=(1.19_{-0.41}^{+0.43})\times 10^{-4}$
with the stellar-density prior, leading to a $\sim 2.9\sigma$ detection.
The mass and radius ratios are translated as
$M_p= 0.18\pm0.07~\Mjup$ and $R_p=0.66\pm0.04~\Rjup$ assuming the stellar
mass and radius. Based on the $\chi^2$ values for the
best-fit parameters, we compare BIC values for the two cases; 
BIC values of $295$ and $285$ are obtained for the eccentric
and circular cases (with the stellar-density prior), respectively, so that fitting with a circular orbit is favored. 
The observed radial velocities and whole light curve are plotted in Figures 
\ref{fig:rv_koi1894} and \ref{fig:phase_koi1894},
with their best-fit models for $e=0$. 
We also show zoomed-in versions of the transit and phase-curve variation
in Figures \ref{fig:transit_koi1894} and \ref{fig:EV_koi1894}, respectively. 
In Figure \ref{fig:EV_koi1894}, the binned flux data used for the fit (250 bins) are plotted
by the red crosses, and black points with errorbars indicate the flux data binned
into 30 bins. 
We also tested several periods around $P_\mathrm{orb}$ and found that about
$P_\mathrm{orb}-4\sigma$ gave the minimum $\chi^2$ value in the global fitting
($\chi^2\sim200$), suggesting that the true period might exist around it.
Nonetheless, the resultant parameters are well consistent with those for the case of
$P_\mathrm{orb}$ within 0.2$\sigma$ level.\footnote{$\Delta T_c$ differs by $\sim 3\sigma$
level.}

Our global analysis indicates the detection of KOI-1894.01 is still marginal, with only 
$\sim 2-3\sigma$ level, and the planet mass seems to be mainly constrained by
the radial velocity data. In order to see if an independent estimate for the planet
mass from the light curve alone gives a comparable result, we fit the folded 
light curve and estimate system parameters without radial velocity data. 
As a result, we obtain $q=(1.27_{-0.62}^{+0.71})\times 10^{-4}$, consistent with
the above result by the global fit. 

%----------------------------------------------------------
%----------------------------------------------------------
%----------------------------------------------------------

\section{Discussion and Summary}\label{summary}
We reported the results of high-precision radial-velocity measurements with
Subaru/HDS for the two {\it Kepler} evolved stars Kepler-91 and KOI-1894, which are
candidates to host transiting planets.

Based on the simultaneous modeling of the radial-velocity data and {\it Kepler}
light curves, we independently confirmed the planetary nature of Kepler-91b.
The radial velocity semi-amplitude of $\sim 55$ m s$^{-1}$ we derived is consistent
with what \citet{barclay:2014} found ($K=67\pm11$ m s$^{-1}$) with
$\sim 1\sigma$, but incompatible with the result by \citet{lillobox:2014b} who reported
$K=93\pm 17$ m s$^{-1}$ with $\gtrsim 2\sigma$, although
the reason for this discrepancy is unknown.
We newly detected a drift of $\sim20$ m s$^{-1}$ between the radial velocities
taken at $\sim$ 1-yr interval thanks to our better measurement precision and
longer period of time of observations compared to the previous ones.
The persistent radial velocity drift is suggestive of a possible presence of another
companion (planet)  outside of Kepler-91b. Due to the lack of data, however, we are
not able to pin down the period nor mass of the companion.
Intensive high-precision radial-velocity monitoring of the star will uncover the unseen companion
and provide hints for formation and evolution of the planetary system.

The estimated parameters for Kepler-91b in Table \ref{tbl:best-fit-Kepler91} are reasonably
in good agreement  with the previous study by \citet{lillobox:2014a, lillobox:2014b}
except for the mass ratio $q$ and scaled semi-major axis $a/R_{\star}$.
This is likely because our radial velocity data show a smaller amplitude and planet
mass is estimated to be small.
Since the flux amplitude due to ellipsoidal variation is approximately proportional
to $q(a/R_{\star})^{-3}$ \citep{shporer:2011}, a small $q$ resulted in the smaller $a/R_{\star}$
in order to  explain the observed flux amplitude. 

The westward phase-shift ($\Delta\phi>0$) of the flux maximum in planetary light 
(from the substellar point) is indicative of the inhomogeneous cloud coverage on the planetary surface. 
This is consistent with the trend that \citet{esteves:2014} found, claiming that 
westward phase-shifts are preferentially seen for relatively cool close-in planets
with equilibrium temperatures of $T_{\rm eq}\lesssim 2500$ K; \citet{lillobox:2014a}
estimated the equilibrium temperature of Kepler-91b to be $\sim 1920-2460$ K, depending
on the assumed heat redistribution parameter. 
The magnitude of the phase-shift ($>0.35$) is, however, unusually large in our
best-fit model compared with the previously reported values for other close-in 
planets \citep{lillobox:2014a}. The reason is unknown, but the huge coverage of 
the planet surface illuminated by the host star\footnote{About $70\%$ of the planet surface is
always illuminated by the host star due to its huge radius.} might be responsible.
Note that we also detected a dip in the folded light curve (Figure \ref{fig:phase_koi2133}) 
around $\phi=0.7$ reported in \citet{lillobox:2014a}, and this dip should have more or less 
affected the fitting result. 

%----------------------------------------------------------
\begin{figure}
\epsscale{1.2}
\plotone{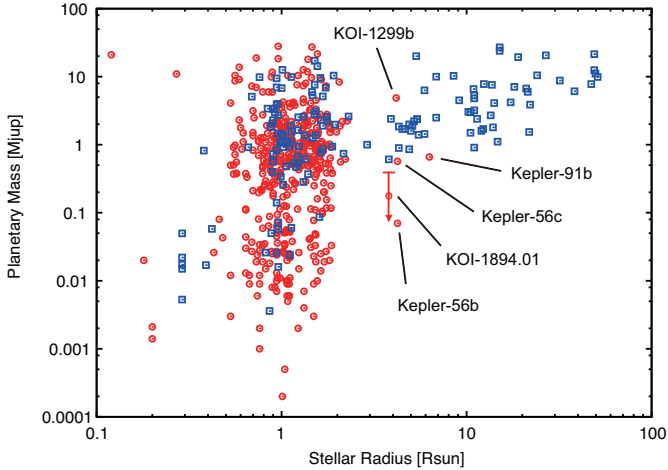}
\caption{Planetary mass plotted against host star's radius. Red circles
and blue squares represent true mass and projected mass, respectively.
Planetary and stellar data were downloaded from NASA Exoplanet Archive, and
only systems whose stellar mass and radius are both reported are plotted.
Median and 3$\sigma$ upper limit value are shown for mass of KOI-1894.01.}
\label{fig:plm-strad}
\end{figure}

%----------------------------------------------------------
\begin{deluxetable*}{lrrrr}
%\tabletypesize{\scriptsize}
\tablecolumns{5}
\tablewidth{0pt}
\tablecaption{Orbital Parameters for KOI-1894}
\tablehead{
\colhead{Parameter}& \multicolumn{4}{c}{Value}\\
 & \colhead{$e=0$ with $\rho_{\star}$ prior} & \colhead{$e$ free with $\rho_{\star}$ prior} & \colhead{$e=0$ without $\rho_{\star}$ prior} & \colhead{$e$ free without $\rho_{\star}$ prior}}
\startdata
\label{tbl:best-fit-KOI1894}
$a/R_{\star}$ & 3.79$^{+0.24}_{-0.26}$ & 3.68$^{+0.27}_{-0.31}$ & 4.25$^{+0.37}_{-0.60}$ & 3.24$^{+0.49}_{-0.57}$ \\
$b$ & 0.634$^{+0.065}_{-0.073}$& 0.56$^{+0.12}_{-0.10}$ & 0.48$^{+0.19}_{-0.21}$ & 0.68$^{+0.12}_{-0.095}$\\
$u_1+u_2$ & 0.724$^{+0.0099}_{-0.010}$& 0.723$^{+0.011}_{-0.0098}$  & 0.724$^{+0.010}_{-0.010}$& 0.7271$^{+0.0079}_{-0.0096}$\\
$u_1-u_2$ & 0.391$^{+0.0098}_{-0.010}$  & 0.3908$^{+0.0097}_{-0.0097}$ & 0.392$^{+0.010}_{-0.0099}$  & 0.391$^{+0.010}_{-0.0096}$\\
$q(=M_p/M_{\star})$ (10$^{-4}$) & 1.19$^{+0.43}_{-0.41}$ & 1.02$^{+0.44}_{-0.49}$ & 1.29$^{+0.50}_{-0.51}$ & 0.70$^{+0.47}_{-0.40}$\\
$F_0$ (10$^{-5}$) & 0 (fixed) & 0 (fixed) & 0 (fixed) & 0 (fixed)\\
$F_1$ (10$^{-5}$) & 0 (fixed) & 0 (fixed) & 0 (fixed) & 0 (fixed)\\
$\Delta\phi$ & 0 (fixed) & 0 (fixed) & 0 (fixed) & 0 (fixed)\\
$R_p/R_{\star}$ & 0.01777$^{+0.00041}_{-0.00035}$  & 0.01739$^{+0.00063}_{-0.00035}$ & 0.01717$^{+0.00081}_{-0.00042}$  & 0.01790$^{+0.0010}_{-0.00046}$\\
$e\cos\omega$ & 0 (fixed)& 0.052$^{+0.12}_{-0.077}$ & 0 (fixed)& 0.047$^{+0.11}_{-0.23}$\\
$e\sin\omega$ & 0 (fixed)  & 0.085$^{+0.086}_{-0.095}$ & 0 (fixed)  & 0.10$^{+0.14}_{-0.13}$\\
$\Delta T_c$ (10$^{-3}$ day) & $-$0.25$^{+0.25}_{-0.40}$& $-$0.53$^{+0.51}_{-0.79}$ & $-$0.24$^{+0.25}_{-0.41}$& $-$0.57$^{+2.0}_{-1.0}$\\
$\gamma$ (m s$^{-1}$) & $-$10.6$^{+4.1}_{-4.1}$ & $-$10.0$^{+4.8}_{-3.9}$ & $-$11.3$^{+4.8}_{-4.7}$ & $-$6.8$^{+3.7}_{-4.9}$\\
$\chi^2$ & 234 & 233 & 235 & 233\\
BIC &  285 & 295 & 286 & 294 \\
\hline
$i_o$ (deg) & 80.4$^{+1.6}_{-1.8}$ & 80.2$^{+2.1}_{-2.6}$ & 83.5$^{+3.1}_{-4.1}$ & 76.2$^{+4.0}_{-10.1}$\\
$e$ & 0 (fixed) & 0.149$^{+0.088}_{-0.077}$  & 0 (fixed) & 0.186$^{+0.15}_{-0.097}$ \\
$\omega$ (deg) & 0 (fixed) & 61$^{+45}_{-61}$ & 0 (fixed) & 65$^{+74}_{-70}$\\
$M_p$ ($\Mjup$) & 0.18$\pm$0.07 & 0.15$^{+0.07}_{-0.08}$ & 0.19$\pm$0.08 & 0.10$^{+0.07}_{-0.06}$\\
$R_p$ ($\Rjup$) & 0.66$\pm$0.04 & 0.64$^{+0.04}_{-0.03}$ & 0.63$\pm0.04$ & 0.66$^{+0.05}_{-0.04}$
\enddata
\end{deluxetable*}

As for KOI-1894, we did not detect any statistically significant radial-velocity variations
with our measurement precision of ~9--15 m s$^{-1}$ and the number of data points.
We excluded the possibility of a grazing transit by a binary companion for the star
and set an upper limit on the mass of KOI-1894.01 to be $0.39~\Mjup$ by our
non-detection of radial-velocity variations with 3$\sigma$ level.
However, we detected possible radial-velocity variations with a semi-amplitude
of $\sim$15 m s$^{-1}$ in phase with ellipsoidal variations of the star with 2--3$\sigma$
level. Although we can not say the detection is statistically significant
at this stage, it suggests that the KOI-1894.01 could be a sub-saturn-mass planet.
Figure \ref{fig:plm-strad} shows distribution of mass of exoplanets currently known
plotted against their host star's radius. As seen in the figure, KOI-1894.01 could be
one of the lowest mass planets ever discovered around evolved stars together
with Kepler-56b, a super-neptune-mass planet ($M_p=0.07~\Mjup$) detected
via TTV (transit timing variation) method \citep{huber:2013b}.\footnote{Kepler-56b
is the inner planet of a double planetary system with inclined orbital plane
\citep{huber:2013b}.}
Actually the stellar parameters for KOI-1891 are similar to those of Kepler-56
\citep[$M_{\star}=1.32~\Msun$, $R_{\star}=4.23~\Rsun$]{huber:2013a}.
Confirmation of KOI-1894.01 is highly encouraged in order to uncover such a new
population of sub-saturn and super-neptune planets around relatively massive evolved
stars, which have rarely been found so far either by radial-velocity surveys or
transit ones.

%% Included in this acknowledgements section are examples of the
%% AASTeX hypertext markup commands. Use \url without the optional [HREF]
%% argument when you want to print the url directly in the text. Otherwise,
%% use either \url or \anchor, with the HREF as the first argument and the
%% text to be printed in the second.

\acknowledgments

This research is based on data collected at Subaru Telescope, which is operated by
National Astronomical Observatory of Japan (NAOJ).
We are grateful to all the staff members of Subaru for their support during the observations. 
This research has made use of the SIMBAD database, operated at
CDS, Strasbourg, France.
T.H. is supported by Japan Society for Promotion of Science
(JSPS) Fellowship for Research (PD:25-3183). 
K.M. is supported by JSPS Research Fellowships for Young Scientists (No. 26-7182) 
and by the Leading Graduate Course for Frontiers of Mathematical
Sciences and Physics.


%----------------------------------------------------------

\end{document}